\documentclass[aps,prc,twocolumn,superscriptaddress]{revtex4}

\usepackage{epsfig}
\begin{document}

%{\hfill \bf  Published in Phys. Rev. C 87, 044325 (2013)} \vspace{1cm}
\begin{minipage}{0.9\textwidth}
\begin{flushright}
\text{\bf Published in Phys. Rev. C 87, 044325 (2013)} 
\end{flushright}
\end{minipage}
\vspace{0.5cm}

%Title of paper
\title{Giant dipole resonance width as a probe for nuclear deformation at finite excitation}

\author{Deepak Pandit}
\email[e-mail:]{deepak.pandit@vecc.gov.in}
\affiliation{Variable Energy Cyclotron Centre, 1/AF-Bidhannagar, Kolkata-700064, India}

\author{Balaram Dey}
\affiliation{Variable Energy Cyclotron Centre, 1/AF-Bidhannagar, Kolkata-700064, India}

\author{Debasish Mondal}
\affiliation{Variable Energy Cyclotron Centre, 1/AF-Bidhannagar, Kolkata-700064, India}

\author{S. Mukhopadhyay}
\affiliation{Variable Energy Cyclotron Centre, 1/AF-Bidhannagar, Kolkata-700064, India}

\author{Surajit Pal}
\affiliation{Variable Energy Cyclotron Centre, 1/AF-Bidhannagar, Kolkata-700064, India}

\author{Srijit Bhattacharya}
\affiliation{Department of Physics, Barasat Govt. College, Barasat, N 24 Pgs, Kolkata - 700124, India }

\author{A. De}
\affiliation{Department of Physics, Raniganj Girls' College, Raniganj-713358, India}

\author{S. R. Banerjee}
\email[e-mail:]{srb@vecc.gov.in}
\affiliation{Variable Energy Cyclotron Centre, 1/AF-Bidhannagar, Kolkata-700064, India}

%Collaboration name if desired (requires use of superscriptaddress
%option in \documentclass). \noaffiliation is required (may also be
%used with the \author command).
%\collaboration can be followed by \email, \homepage, \thanks as well.
%\collaboration{}
%\noaffiliation

\date{\today}

\begin{abstract}
The systematic study of the correlation between the experimental giant dipole resonance (GDR) width and the average deformation $\left\langle \beta \right\rangle$ of the nucleus at finite excitation is presented for the mass region A $\sim$ 59 to 208. We show that the width of the GDR ($\Gamma$) and the quadrupole deformation of the nucleus do not follow a linear relation, as predicted earlier, due to the GDR induced quadrupole moment and the correlation also depends on the mass of the nuclei. The different empirical values of $\left\langle \beta \right\rangle$ extracted from the experimental GDR width match exceptionally well with the thermal shape fluctuation model. As a result, this universal correlation between $\left\langle \beta \right\rangle$ and $\Gamma$ provides a direct experimental probe to determine the nuclear deformation at finite temperature and angular momentum in the entire mass region.

% insert abstract here
\end{abstract}
% insert suggested PACS numbers in braces on next line
% insert suggested keywords - APS authors don't need to do this
%\keywords{}
%\maketitle must follow title, authors, abstract, \pacs, and \keywords
\maketitle

\section{Introduction}

The atomic nucleus is a complex many body quantum system which displays an unbelievably rich and intriguing variety of phenomena. While there are many excitations that are quite irregular and even considered as the manifestation of chaotic motion, the nuclei, on the other hand, also show collective behavior that reflects the dynamical properties of the nuclear system. Giant resonances, the collective mode of excitation, are of particular interest because they currently provide the most reliable information about the bulk behavior of the nuclear many body system. A typical example of this vibrational mode is the isovector giant dipole resonance (IVGDR) in which the neutrons and protons oscillate out of phase against each other \cite{hara01, gaar92}. Interestingly, it is the only giant resonance experimentally studied extensively at finite temperature (T) $\&$ 
angular momentum (J). Consequently, it has become an indispensable tool in nuclear structure physics. 

The GDR decay from the excited nuclei occurs on a time scale that is sufficiently short and thus probes its conditions 
prevailing at that time \cite{gaar92}. The centroid energy of the resonance is inversely proportional to the nuclear radius and provides an idea about the nuclear size. Moreover, the centroid energy is strongly correlated to the nuclear symmetry energy which is a fundamental quantity important for studying the structure of neutron star \cite{tri08}. On the other hand, the width of the resonance corresponds to the damping of this collective vibration due to the viscosity of the neutron and proton fluids \cite{aue75}. Recently, the precise experimental systematics of the GDR widths in hot nuclei have been applied to calculate the ratio of shear viscosity $\eta$ to the entropy volume density $s$ \cite{Dang11}. It is concluded that the ratio $\eta$/$s$  in medium and heavy nuclei decreases with increasing temperature to reach the value (1.3-4)$\times$ $\hbar$/(4$\pi$k$_B$) at T = 5 MeV indicating that nucleons inside a hot nucleus at T = 5 MeV have nearly the same ratio $\eta$/$s$ as that of Quark Gluon Plasma. Microscopically, the total width of the GDR consists of the landau width, the spreading width and the escape width \cite{hara01}. In medium and heavy nuclei, the landau and the escape widths only account for a small fraction and the major contribution of the large resonance width comes from the spreading width \cite{bort01, don01}. Recently, an empirical formula has been derived for the spreading width with only one free parameter by separating the deformation induced widening from the spreading effect \cite{jun08}.	
It is now well known that the GDR strength function splits in the case of a deformed nucleus and the deformation can be estimated from the ratio of the two resonance energies \cite{hara01}. However, for small deformations, the separation is not appreciable and the two resonance energies cannot be identified individually. As a result, the overall width of the GDR increases. Thus, the apparent GDR width ($\Gamma$) can provide us a direct experimental probe to measure the deformation of the atomic nuclei at high temperature and angular momentum. However, this interesting aspect of the apparent GDR width has not been explored so far.

\begin{figure}
\begin{center}
\includegraphics[height=10.0 cm, width=7.5 cm]{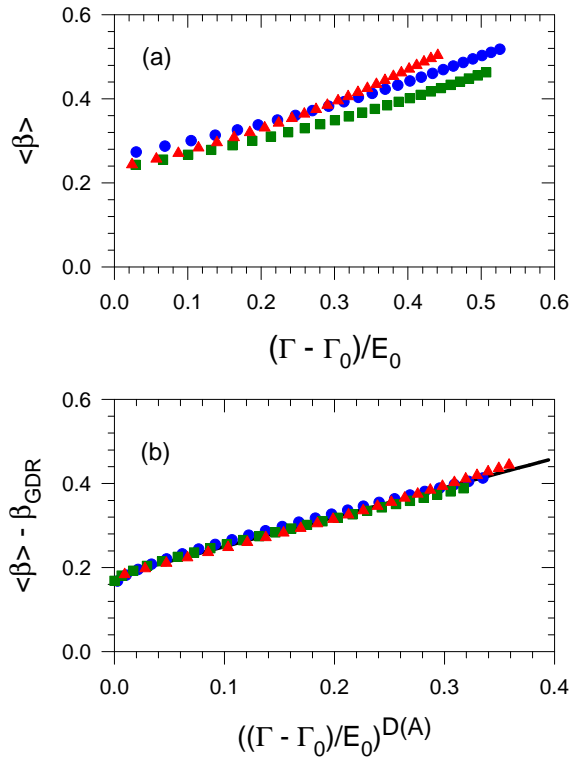}
\caption{\label{Bavg} (color online) Average nuclear deformation vs GDR width for $^{63}$Cu (filled circles), $^{120}$Sn (filled squares) and $^{208}$Pb (filled triangles) for T$\leq$3.5 MeV $\&$ J=15$\hbar$. (a) The average deformation and the GDR width do not follow a linear relationship,
as predicted earlier, when actual T-dependence of the GDR width is taken.
(b) A linear correlation is obtained when $\left\langle \beta \right\rangle$ - $\beta$$_{GDR}$ is plotted as a function (($\Gamma$ - $\Gamma_0$)/$E_0$)$^{D(A)}$. The continuous line is the proposed correlation represented by Eq.\ref{eqn1}}
\end{center}
\end{figure}
	
Several studies of the GDR $\gamma$-decay in hot nuclei have showed that while the GDR centroid energy remains more or less constant with excitation energy, the apparent width of the resonance increases with both temperature and angular momentum \cite{hara01}. The J-dependence of the GDR width is described very successfully within the thermal shape fluctuation model (TSFM) \cite{alh88, alh93, kus98}. As the rotational frequency becomes larger, the nucleus undergoes an oblate flattening due to centrifugal effects. The equilibrium deformation ($\beta$$_{eq}$) increases rapidly with J and, as a consequence, the total GDR strength function undergoes splitting which increases the overall width of the resonance. The model has also been applied successfully to explain the Jacobi shape transition in atomic nuclei \cite{Maj04, Dipu2}. However, it needs to be mentioned that even though the equilibrium deformation of a nucleus increases with J, an increase of GDR width is not evident experimentally until the equilibrium deformation increases sufficiently to affect the thermal average \cite{Cam99}. In particular, as long as  $\beta$$_{eq}$  is less than the variance $\Delta$$\beta$=$\left[\left\langle \beta^2\right\rangle - \left\langle \beta\right\rangle^2\right]^{1/2}$  
the increase of GDR width is not significant. Thus, the competition between $\beta$$_{eq}$ and $\Delta$$\beta$ give rise to the critical angular momentum, observed in all the experiments, below which the GDR width remains nearly constant at its ground state values.	
In the case of T dependence, it is observed that the experimental GDR widths remain more or less constant until T $\sim$ 1 MeV and increase thereafter with T. The increase of the GDR width above T = 1.5 MeV can be explained reasonably well within the TSFM. The model proposes that the nucleus does not posses a single well-defined shape but rather explores a broad ensemble of mostly quadrupole shapes because of thermal fluctuation around an equilibrium shape. Thus, in adiabatic assumption i.e. when the shape fluctuations are slow compared to the frequency shift, the observed GDR width then results from a weighted average over all frequencies associated with the possible shapes  \cite{alh88, Orm96}. This gives rise to T-driven broadening of the width. However, the model fails to explain the experimental data below T = 1.5 MeV in different mass regions  \cite{supm11, heck03, cam03, dipu10, dipu12}. Recently, it has been shown that the GDR vibration itself produces a quadrupole moment causing the nuclear shape to fluctuate even at T=0 MeV \cite{Sim03, Sim09, dipu12}. Therefore, when the giant dipole vibration, having its own intrinsic fluctuation is used as a probe to view the thermal shape fluctuations, it is unlikely to feel the thermal fluctuations that are smaller than its own intrinsic fluctuation. The discrepancy between the experimental data and the TSFM predictions at low T is attributed to the competition between the GDR induced fluctuation ($\beta$$_{GDR}$) and the variance of the deformation $\Delta\beta$  due to thermal fluctuations. This gives rise to a critical temperature ($T_c$) in the increase of the GDR width. A new phenomenological model has been proposed by invoking this idea and is called critical temperature included fluctuation model (CTFM) \cite{dipu12}. The model gives an excellent description of the GDR width for both T $\&$ J in the entire mass range.

In this paper, we present a systematic study of a universal correlation between the giant dipole resonance width and the average deformation of the nuclei. We show that the relationship between the GDR width and $\left\langle \beta \right\rangle$ is non-linear 
because of the GDR induced quadrupole moment. We also find good agreement between  $\left\langle \beta \right\rangle$ extracted from the experimental GDR width and the TSFM calculation for both T $\&$ J in the mass region A $\sim$ 59 to 208. 

\begin{figure}
\begin{center}
\includegraphics[height=13.0 cm, width=7.5 cm]{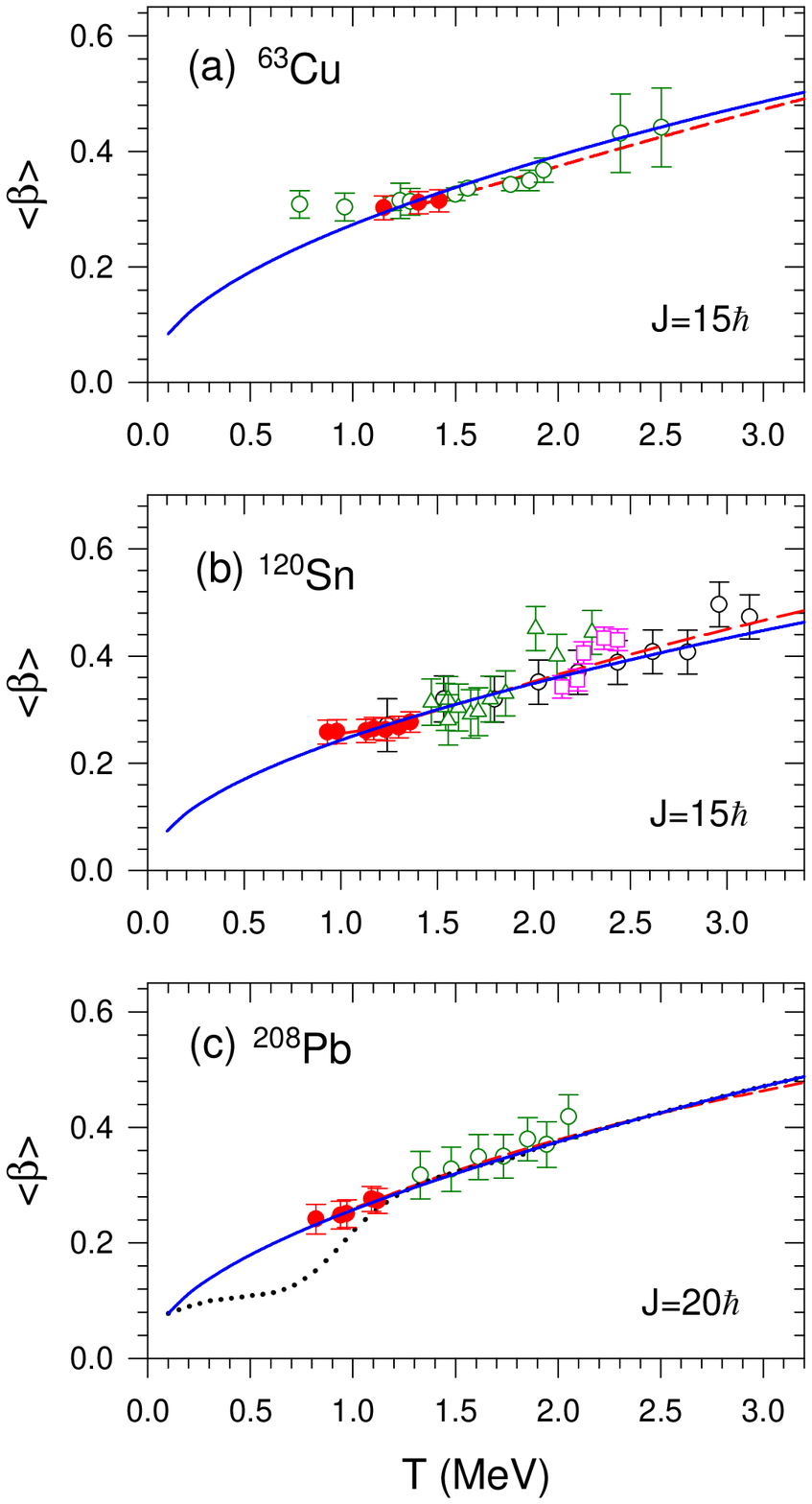}
\caption{\label{B_T} (color online) The average deformation as a function of T for
$^{63}$Cu, $^{120}$Sn and $^{208}$Pb. (a) The filled circles are data 
from Refs\cite{dipu12} while  open circles are from Refs\cite{kus98, kic87} for $^{63}$Cu.
(b) $^{120}$Sn data (open circles \cite{Bau98}, open squares \cite{kel99}, up triangle \cite{kus98}) 
are shown along with $^{119}$Sb data (filled circles \cite{supm11}).  
(c) $^{208}$Pb data (open circles \cite{Bau98}) along with $^{201}$Tl data (filled circles \cite{dipu12}).
The continuous lines correspond to the TSFM calculations while the dashed lines represent the average deformations
estimated using the GDR width from CTFM. The dotted line in (c) is the average deformation calculated for $^{208}$Pb
including the shell effect.}
\end{center}
\end{figure}
\begin{figure}
\begin{center}
\includegraphics[height=13.0 cm, width=7.5 cm]{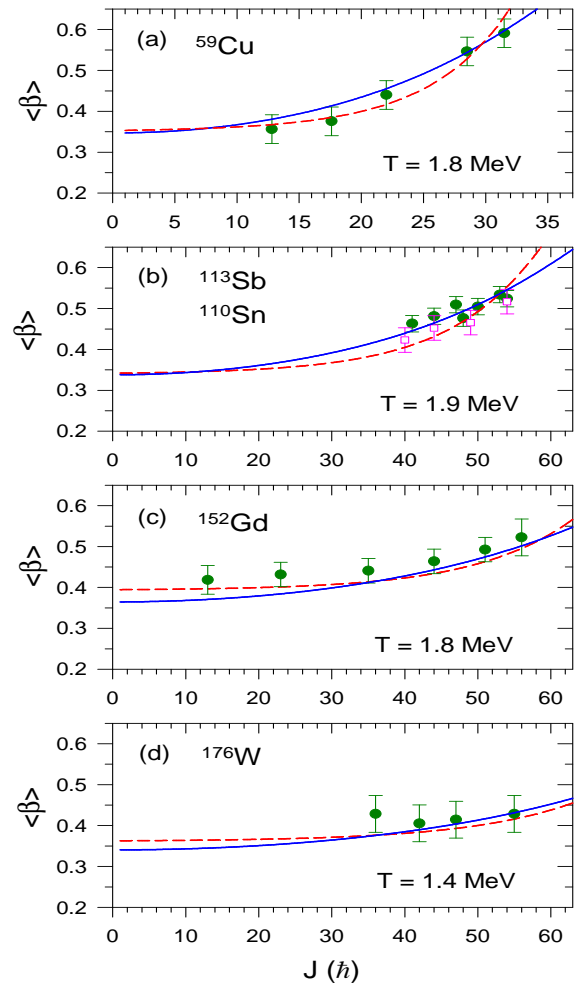}
\caption{\label{B_J} (color online) The average deformation as a function of J. 
(a) The filled cirlces represent the emperical deformation for $^{59}$Cu. 
(b) The filled cirlces correspond to $^{113}$Sb data while the open squares represent the $^{110}$Sn data. 
(c) The filled cirlces represent the data for $^{152}$Gd.
(d) The filled cirlces represent the data for $^{176}$W.
The continuous lines in (a), (b), (c) and (d) correspond to the TSFM calculations while the dashed lines represent the average deformations
estimated using the GDR width from CTFM.}
\end{center}
\end{figure}

\section{The formalism and the universal correlation}

It is very interesting to note that the increase of the GDR width with both J and T can be explained by the spreading due to the shift in the centroid
energy of the GDR modes for each deformation. Moreover, simple scaling functions depending only on J, T and A represent the experimental data on the GDR width remarkably well. Hence, in principle, there should exist a correlation between the width of the GDR and the average deformation $\left\langle \beta \right\rangle$ of the nucleus at finite T and J. 
The mass dependence of the mean value of $\beta$ and FWHM of the GDR width was studied by Mattiuzzi et al., \cite{Mat97} and hinted towards the correlation between $\left\langle \beta \right\rangle$ and GDR width. It was shown that the oblate flattening due to the angular momentum would be small for heavier nuclei and thus the FWHM should exhibit less dependence on angular momentum. 
A few years later, it was shown within the TSFM  \cite{kus03} that $\left\langle \beta \right\rangle$ is directly correlated to the quantity 
($\Gamma(J,T,A)$ - $\Gamma_0$)/$E_0$ where $\Gamma_0$ and $E_0$ represent the width and centroid energy of the GDR for a spherical nucleus, respectively. However, the comparison was made only for Sn nucleus and the ansatz failed \cite{kus03} to represent the temperature dependence of $\left\langle \beta \right\rangle$ deduced from experimental data with the TSFM calculation. 

We remark here that the TSFM  does not represent the proper T-dependence of the GDR width as it does not take into account the fluctuations introduced by the GDR motion \cite{dipu12}. Hence, the linear relationship, proposed earlier, failed to explain the T-dependence of $\left\langle \beta \right\rangle$  derived from the experimental GDR width with the TSFM calculation \cite{kus03}. In Figure \ref{Bavg}(a), we plot $\left\langle \beta \right\rangle$  as a function of ($\Gamma$ - $\Gamma_0$)/$E_0$ for $^{63}$Cu, $^{120}$Sn and $^{208}$Pb as systematic data exist in this mass region over a wide range of T. In this case, the GDR width was derived from CTFM which represents precise T-dependence of the GDR width while $\left\langle \beta \right\rangle$ was calculated under the TSFM framework using the boltzmann probability $e^{-F(\beta, \gamma)/T}$ with the volume element $\beta^{4} sin(3\gamma) d\beta d\gamma$ described in Ref\cite{Dipu2}. It could be clearly seen from Figure \ref{Bavg}(a) that different nuclei have different slopes as well as different intercepts when actual T-dependence of the GDR width is taken into account. We remark here that the width of the GDR and  $\left\langle \beta \right\rangle$  of the nucleus cannot be directly compared as GDR vibration itself produces a fluctuation and cannot probe the variation that are smaller than its own intrinsic fluctuation \cite{dipu12}. In fact,  $\left\langle \beta \right\rangle$  should be correlated to the width of the GDR along with the deformation induced by GDR motion ($\beta$$_{GDR}$). Interestingly, a linear correlation is indeed obtained 
when $\left\langle \beta \right\rangle$ - $\beta$$_{GDR}$ is plotted as a function of (($\Gamma$ - $\Gamma_0$)/$E_0$)$^{D(A)}$ where $D(A)$ has a small mass dependence (Figure \ref{Bavg}(b)). We propose the correlation between the average deformation of the nucleus and the width of the GDR as
\begin{equation}
\beta_{exp} = 0.18 + \beta_{GDR} + 0.7 \left(\frac{\Gamma(J,T,A) - \Gamma_0}{E_0}\right)^{D(A)} \label{eqn1}
\end{equation}
where, 
\begin{eqnarray}
\beta_{GDR}= 0.04 + 4.13/A \nonumber
\end{eqnarray}
\begin{eqnarray}
D(A) = 2 - 0.0036A. \nonumber
\end{eqnarray}

\section{Results and Discussions}

It is interesting to note that the power coefficient 
D(A) decreases with an increase in mass, as found previously for both $T_c$ and $\beta$$_{GDR}$ \cite{dipu12}.  In order to verify the correlation, the experimental data of $^{63}$Cu\cite{kus98, kic87, dipu12}, $^{119}$Sb\cite{supm11}, $^{120}$Sn\cite{kel99, kus98, Bau98}, $^{201}$Tl\cite{dipu12} and $^{208}$Pb\cite{Bau98} were used to extract the empirical deformation using  Eq.\ref{eqn1}. The values of $\left\langle \beta \right\rangle$ extracted from the experimental data are directly compared with the TSFM calculation (continuous line) in Figure \ref{B_T}. As can be seen, in all the three mass regions, there is an excellent match between the experimental data and the TSFM calculation. The empirical deformations as a function of angular momenta for the nuclei $^{59}$Cu\cite{Dre95}, $^{110}$Sn\cite{Bra04}, $^{113}$Sb\cite{Sri08}, $^{152}$Gd\cite{Drc10} and $^{176}$W\cite{Mat95} are also compared with the TSFM calculations in Figure \ref{B_J}. 
Interestingly, in this case too the experimental data and TSFM calculations are in good agreement in the entire mass range. 
The average deformations estimated using the GDR widths predicted by phenomenological CTFM have also been compared with the experimental data 
and found to match reasonably well (dashed lines in Figure \ref{B_T} $\&$ Figure \ref{B_J}). 
For all the nuclei, the centroid energy of the GDR was calculated using the systematic $E_{GDR} = 31.2A^{-1/3} + 20.6A^{-1/6}$ \cite{hara01}.
For the calculation of the width of the spherical nucleus we used the relation $\Gamma$$_0$=0.05E$^{1.6}_{GDR}$ that has been derived recently by disentangling the effects of spreading width and the deformation induced widening \cite{jun08}. 

We mention here that the experimental GDR width can only be applied above the critical temperature ($T_c$ = 0.7 + 37.5/A) to measure the nuclear deformation at finite excitation energy. Below $T_c$, the GDR vibration does not view the thermal fluctuations, as they are smaller than its own intrinsic GDR fluctuation \cite{dipu12}. Consequently, the experimental data and TSFM are not in good agreement below $T_c$ (Figure \ref{B_T}). It is shown in Ref\cite{dipu12} that shell effects indeed play an important role at low temperature in A $\sim$ 200 as it increases the critical temperature from 0.5 MeV to $\sim$ 0.9 MeV. The value of $\left\langle \beta \right\rangle$ for $^{208}$Pb was also calculated including shell effects (dotted line in Figure \ref{B_T}) and compared with the experimental data. It can be seen that the experimental data and TSFM match very well above 0.9 MeV as shell effects have already reduced by a factor of 10. Hence, the shell effects are directly included in the definition of critical temperature. In general, Eq.\ref{eqn1} can be applied above $T_c$ as well as below the Jacobi transition point ($J_c$ $\sim$ 1.2A$^{5/6}$ \cite{kus03}) since $\gamma_{min}$ displays an abrupt change from $\gamma$=$\pi$/3 to 0 and may not follow the simple correlation. Nevertheless, the empirical data and TSFM match exceptionally well above $T_c$ in the entire mass region for all values of T $\&$ J. The GDR width, therefore, provides us a direct experimental probe to assess the nuclear deformation at finite temperature and angular momentum. Moreover, this novel idea of universal correlation should provide new insights into the modification of the TSFM at low temperature by calculating the effective deformation probed by the GDR motion at finite excitation energy i.e. the quadrupole moment induced by the GDR motion should be included in the TSFM. 

\section{Summary and Conclusion}

In summary, we have presented a systematic study of the correlation between the width of the giant dipole resonance and the quadrupole deformation of the nucleus at finite excitation. We have shown, by comparing $\left\langle \beta \right\rangle$ calculated within TSFM and GDR width $\Gamma$ estimated through CTFM, that the correlation between $\left\langle \beta \right\rangle$ and $\Gamma$ is non-linear due to the GDR induced quadrupole moment. The different experimental data, extracted using the proposed correlation, are in good agreement with the TSFM calculation in the entire mass range. Consequently, the apparent GDR width can be used as a direct experimental probe to measure the nuclear deformation as a function of T $\&$ J where the different GDR resonance energies due to deformation cannot be separated.


\begin{thebibliography}{99}

\bibitem{hara01} M. N. Harakeh and A. van der Woude, Giant Resonances, Fundamental High-frequency Modes of
Nuclear Excitation, Clarendon Press, Oxford, 2001.
\bibitem{gaar92} J. J. Gaardhoje, Ann. Rev. Nucl. Part. Sci. {\bf 42}, 483 (1992).
\bibitem{tri08} Luca Trippa, Gianluca Col`o, and Enrico Vigezzi, Phys. Rev. C {\bf 77}, 061304(R) (2008).
\bibitem{aue75} N. Auerbach and A. Yeverechyahu, Annals of Physics  {\bf 95}, 35 (1975).
\bibitem{Dang11} Nguyen Dinh Dang, Phys. Rev. C {\bf 84},  034309 (2011).
\bibitem{bort01} P. F. Bortignon et al, Nucl. Phys. A 460, 149 (1986).
\bibitem{don01} P. Donati et al, Phys. Lett. B 383, 15 (1996).
\bibitem{jun08}  A. R. Junghans et al, Phys. Lett. B 670, 200 (2008).
\bibitem{alh88} Y. Alhassid, B. Bush and S .Levit, Phys. Rev. Lett. 61, 1926 (1988).
\bibitem{alh93} Y. Alhassid and N. Whelan Nucl. Phys. A565, 427 (1993).
\bibitem{kus98}  Dimitri Kusnezov et al., Phys. Rev. Lett. {\bf 81},  542 (1998).
\bibitem{Maj04} A. Maj et al., Nucl. Phys. A 731, 319 (2004).
\bibitem{Dipu2} Deepak Pandit et al, Phys. Rev. C  81, 061302(R) (2010).
\bibitem{Cam99}	 F. Camera et al., Nucl. Phys. {\bf A649},  115 (1999).
\bibitem{Orm96}  W. E. Ormand et al., Phys. Rev. Lett. {\bf 77}, 607 (1996).
\bibitem{supm11} S. Mukhopadhayay et al., Phys. Lett. B {\bf 709}, 9 (2012).
\bibitem{heck03} P. Heckman et al., Phys. Lett. B {\bf 555},) 43 (2003.
\bibitem{cam03}  F. Camera et al., Phys. Lett. B {\bf 560}, 155 (2003).
\bibitem{dipu10}  Deepak Pandit et al., Phys. Lett. B {\bf 690}, 473 (2010).
\bibitem{dipu12}  Deepak Pandit et al., Phys. Lett. B {\bf 713}, 434 (2012).
\bibitem{Sim03}  C. Simenel et al., Phys. Rev. C {\bf 68}, 024302 (2003).
\bibitem{Sim09}	 C. Simenel et. al., Phys. Rev. C {\bf 80}, 064309 (2009).
\bibitem{Mat97}  M. Mattiuzzi et al.,Nucl. Phys. {\bf A612}, 262 (1997).
\bibitem{kus03} Dimitri Kusnezov and W. Erich Ormand, Phys. Rev. Lett. {\bf 90}, 042501 (2003).
\bibitem{kic87}  M. Kici\'{n}ska-Habior et al., Phys. Rev. C {\bf 36}, 612 (1987).
\bibitem{kel99}  M. P. Kelly et al., Phys. Rev. Lett. {\bf 82}, 3404 (1999).
\bibitem{Bau98}  T. Baumann et al., Nucl. Phys. {\bf A635}, 248 (1998).
\bibitem{Dre95}  Z. M. Drebi et al., Phys. Rev. C {\bf 52},  578 (1995). 
\bibitem{Bra04}  A. Bracco et al., Phys. Rev. Lett. {\bf 74},  3748 (1995).
\bibitem{Sri08}  Srijit Bhattacharya et al., Phys. Rev. C {\bf 77}, 024318 (2008).
\bibitem{Drc10}  D. R. Chakrabarty et al., J. Phys. G: Nucl. Part. Phys. {\bf 37}, 055105 (2010). 
\bibitem{Mat95}  M. Mattiuzzi et al., Phys. Lett. B {\bf 364}, 13 (1995).



%\bibliography{basename of .bib file}
\end{thebibliography}
\end{document}